\documentclass[pra,12pt,onecolumn,preprintnumbers,amsmath,amssymb,superscriptaddress,longbibliography]{revtex4-1}
\newcommand{\br}{{\bf r}}

\newcommand{\bk}{{\bf k}}

\usepackage{graphicx, epsf}
\usepackage{subfigure}
\usepackage{color}
\usepackage{braket}
\usepackage{bm}
\usepackage{float}

\begin{document}
\title{Do the surface Fermi arcs in Weyl semimetals survive disorder?}

\author{Justin H. Wilson}
\affiliation{Institute of Quantum Information and Matter and Department of Physics, California Institute of Technology, Pasadena, CA 91125 USA}
\author{J. H. Pixley}
\affiliation{Condensed Matter Theory Center, Department of Physics, University of Maryland, College Park, Maryland 20742- 4111 USA}
\affiliation{Department of Physics and Astronomy, Center for Materials Theory, Rutgers University, Piscataway, NJ 08854 USA}
\author{David A. Huse}
\affiliation{Physics Department, Princeton University, Princeton, NJ 08544, USA}
\author{Gil Refael}
\affiliation{Institute of Quantum Information and Matter and Department of Physics, California Institute of Technology, Pasadena, CA 91125 USA}
\author{S. Das Sarma}
\affiliation{Condensed Matter Theory Center, Department of Physics, University of Maryland, College Park, Maryland 20742- 4111 USA}

\date{\today}

\begin{abstract}
We theoretically study the topological robustness of the surface physics induced by Weyl Fermi-arc surface states in the presence of short-ranged quenched disorder and surface-bulk hybridization.
This is investigated with numerically exact calculations on a lattice model exhibiting Weyl Fermi-arcs.
We find that the Fermi-arc surface states, in addition to having a finite lifetime from disorder broadening, hybridize with nonperturbative bulk rare states making them no longer bound to the surface (i.e.\ they lose their purely surface spectral character). 
Thus, we provide strong numerical evidence that the Weyl Fermi-arcs are not topologically protected from disorder.
Nonetheless, the surface chiral velocity is robust and survives in the presence of strong disorder, persisting all the way to the Anderson-localized phase by forming localized current loops that live within the localization length of the surface.
Thus, the Weyl semimetal is not topologically robust to the presence of disorder, but the surface chiral velocity is. 
\end{abstract}

\maketitle

\begin{figure}
\includegraphics{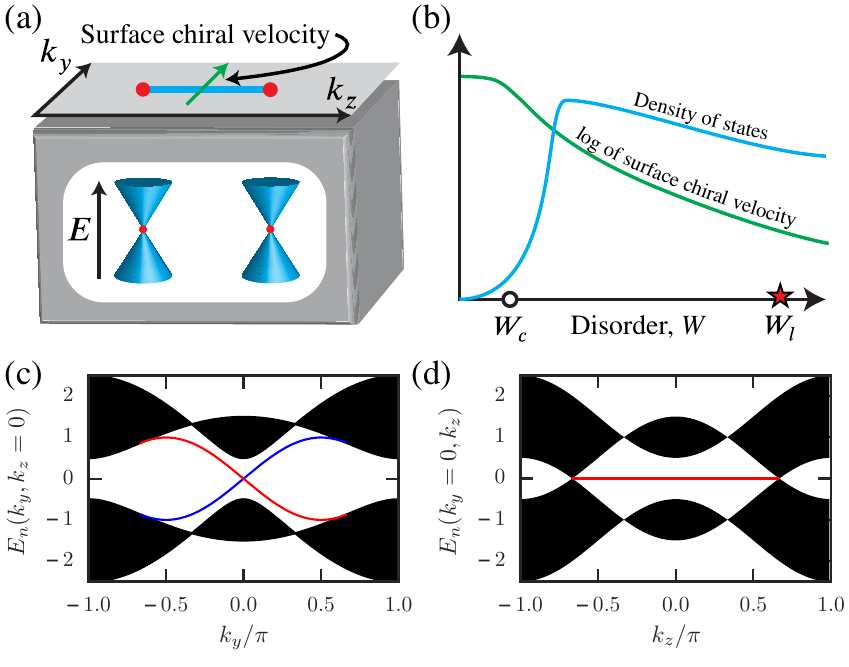}
\caption{(a) Schematic of a Weyl semimetal with two cones in the bulk and therefore a single Fermi arc. The chiral velocity is defined perpendicular to the arc. (b) Schematically, how the density of states and (log of) the surface chiral velocity change in the phases and regimes this model exhibits (horizontal-axis). Setting $m=(3/2)t$, both quantities are evaluated as disorder averages at $E=0$. The diffusive metal phase lies within $0<W<W_l$ and the Anderson insulator for $W>W_l$. (c,d) For $m=(3/2)t$, we plot cuts of the dispersion for Eq.~\eqref{eqn:ham} in the clean limit with open boundary conditions displaying the bulk bands and the topological surface Fermi arc states (red,blue) dispersing like $E(k_y,k_z) = \pm t\sin(k_y)$ in the pseudogap. (c) shows $E(k_y,k_z)$ versus $k_y$ with $k_z=0$, and (d) is  $E(k_y,k_z)$ versus $k_z$ with $k_y=0$; Weyl points at ${\bf K}_W=(0,0,\pm2 \pi/3)$ can be seen.
}
\label{fig:dispersion-cuts}
\end{figure}

Weyl semimetals have recently been experimentally discovered in weakly correlated zero gap semiconductors such as TaAs~\cite{Xu2015,Lv2015,Inoue1184}, NbAs \cite{xu2015discovery}, and TaP \cite{xu2016observation} as well as the strongly correlated material Mn$_3$Sn \cite{kuroda2017evidence}.
Thus, Weyl semimetals (WSMs) are now included in the growing tapestry of topological materials \cite{Hosur2013RecentSemimetals,ArmitageWeyl}.
These gapless three-dimensional materials have nodes in the momentum-space band structure which provide sources and sinks of Berry flux that lead to a set of topological surface states, and the set of these states at a given Fermi energy and restricted to a single surface constitute a ``Fermi arc.'' Fermi arcs begin and end at the projection of the bulk Fermi surface, as seen in Fig.~\ref{fig:dispersion-cuts}(a).
However, with a gapless bulk spectrum, it is not clear how robust these surface states are.

Surface states are a hallmark of topological physics, the pure manifestation of the bulk-boundary correspondence \cite{Jackiw1976Solitons1/2}. 
When the bulk possesses an energy gap at the Fermi energy, topological edge modes are robust to small perturbations \cite{Hasan2010}  and can seem to violate various no-go theorems.
In topological superconductors, the edge can host bound Majorana fermions \cite{Kitaev2001UnpairedWires,Fu2008}, while
quantum Hall edge states host a single chirality \cite{Laughlin1981,Stormer1999}, and three-dimensional topological insulators (TIs) host an odd number of Dirac cones on each surface \cite{Hasan2010}.
The protection and anomalous properties of these edge states make them ideal for high-performance electronics \cite{Konig2007QuantumWells,Chang2013ExperimentalInsulator} and as the building blocks of a quantum computer \cite{Fu2008,Sau2010,Nayak2008,DasSarma2015MajoranaComputation}.
Since surface Fermi arcs represent the bulk-boundary correspondence in WSMs, understanding their robustness (or not) in the presence of disorder is crucial.

However, topological protection is thrown into question for WSMs.
Effects of disorder in the bulk of Weyl (and Dirac) semimetals has been well studied \cite{Fradkin1986,*Fradkin1986a,goswami_quantum_2011,kobayashi_density_2014,sbierski_quantum_2014,Nandkishore2014,Pixley2015,syzranov_critical_2015,altland_effective_2015,chen_disorder_2015,liu_effect_2016,shapourian_phase_2016,Gorbar2016,Pixley2016,Pixley2016a,Pixley2017,JHWilson2017pip,Holder2017,Gurarie2017TheoryDimensions,Syzranov2017High-DimensionalSystems}.
Recently, approximate instanton calculations \cite{Nandkishore2014} and exact numerics \cite{Pixley2016,Pixley2016a,Pixley2017,JHWilson2017pip} conclusively find that non-perturbative 
rare region effects drive WSMs into a diffusive metal phase for any non-zero disorder despite earlier work, based on mean field and perturbative RG theories, erroneously finding a phase transition from semimetal to diffusive metal at finite disorder \cite{Fradkin1986,*Fradkin1986a,Syzranov2017High-DimensionalSystems,ArmitageWeyl}.
These rare region effects, not accessible in mean field theory or perturbative RG theories, round out the semimetal-to-metal transition into a cross-over dubbed an avoided quantum critical point (AQCP) \cite{Pixley2016}.
It is therefore a natural question, and the subject of this article, to determine the robustness of the surface states in the presence of disorder, given that the bulk Weyl semimetal phase is destroyed by any finite disorder.
The consequences of disorder on WSM topology and correspondingly on the Fermi arc surface states is a matter of great importance from the dual perspectives of fundamental principle and practical applications.

For weak TIs, disorder breaks the symmetry responsible for topological protection, but nonetheless the surface states remain \cite{Ringel2012}, and when disorder closes the gap in a strong TI, a remnant of the edge is still preserved \cite{Meyer2013}.
Previous work on WSMs and Chern insulators has suggested that a finite Hall conductivity~\cite{chen_disorder_2015,shapourian_phase_2016} and surface transport~\cite{liu_effect_2016} persist for finite disorder even well into the metallic phase.
However, while both TIs and WSMs have non-perturbative rare states, only the bulk WSM is destroyed by them.
In weakly disordered TIs, rare Lifshitz states populate the bulk band gap \cite{Mieghem1992,Kramer1993};
they are exponentially localized (with no level repulsion) and dilute enough to not couple the surfaces, i.e. the bulk gap provides topological protection to disorder. 
On the other hand, in WSMs the rare states are power-law quasi-localized (with non-zero level repulsion) and ``fill in'' the pseudogap; this gives the Weyl quasiparticles a finite lifetime \cite{Nandkishore2014,Pixley2017,Gurarie2017TheoryDimensions} and a finite DC conductivity \cite{Nandkishore2014,Holder2017}.
Therefore, 
In this sense, it is unclear how the surface states in WSMs might survive the presence of a (weak) random potential.

\begin{figure*}[t!]
    \centering
    \includegraphics{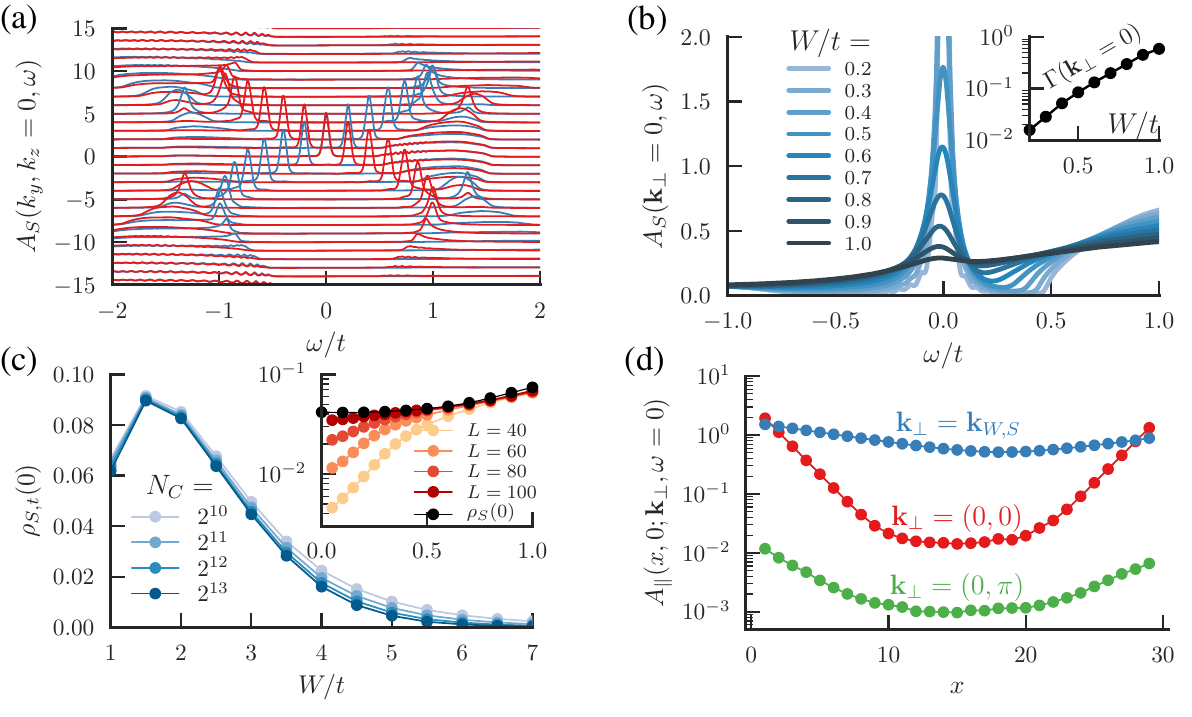}
    \caption{(a) Surface electronic dispersion curves (EDCs) where each value of $A_S(k_y,k_z,\omega)$ for $k_z=0$ with $k=k_y$ is shifted by $(L/2\pi)k$; blue is the top surface and bottom is red. (b) The spectral function versus $\omega$ on the surface at $\bk_{\perp}=0$ for various disorder strengths. We see a smooth broadening of the Fermi arc peak with disorder, as captured by the width of the spectral function $\Gamma(\mathbf k_\perp = 0)$ shown in the inset.
(c) The typical DOS on the surface for strong disorder and weak disorder (inset). As the system size $L$ increases, the typical DOS on the surface converges to the average surface DOS implying the arcs do not localize for weak disorder; for strong disorder, we find the bulk and surface localization transitions agree.
(d) Average spectral weight for states of definite momentum on the surface to tunnel into the bulk for three representative surface momenta on the arc: $\bk_{\perp}=0$, at the Weyl node projection $\bk_{W,S}=(0,2\pi/3)$, and off the arc $\bk_{\perp}=(0,\pi)$, all computed at $W=0.5t$ and $L=30$. The finite value of the spectral weight in the middle of the sample indicates surface-bulk hybridization.
}
    \label{fig:GreensfunctionTypicalDOS}
\end{figure*}

After first principles band structure calculations suggested the existence of WSMs \cite{Wan2011,Weng2015,huang2015weyl}, the Weyl Fermi-arc surface states were observed in photoemission~\cite{Xu2015,Lv2015,xu2015discovery,xu2016observation}  and scanning tunneling microscopy \cite{Inoue1184} experiments on relatively clean materials. 
This makes our central question important for the development of potential technological applications of the surface states, e.g.\ as a ``catalyst'' in solar cells \cite{Rajamathi2016WeylCatalysts}. 
Do the Weyl Fermi-arcs (or any remnant of them) survive disorder?

In this work, we study the effects of short-ranged disorder on Weyl Fermi-arcs numerically in a cubic lattice model that represents a time-reversal broken Weyl semimetal. 
Using kernel polynomial method (KPM), Lanczos, and exact diagonalization, we compute various properties of the arcs.
We establish that the surface only localizes when the bulk becomes an Anderson insulator \cite{Pixley2015}. 
We also establish that non-perturbative quasilocalized rare bulk states hybridize with surface states, giving the arcs spectral weight in the bulk, thereby concluding that the Weyl Fermi-arcs are not topologically protected against even weak disorder. 
Nonetheless, we show that the surface chiral velocity persists deep into the diffusive metal regime, which establishes one aspect of the Fermi arcs displaying a remarkable stability. 
Thus, spectroscopic measurements will continue to see a Fermi arc even in the presence of disorder although this is no longer a protected surface state.
Unexpectedly, the surface chiral velocity survives even in the Anderson insulating phase by inducing local current loops into the bulk that live within the localization length of the surface but cannot contribute to conductivity.
% All data and code used here are available upon request.

\section*{Model and clean surface states}
The tight-binding model used is \cite{JHWilson2017pip}
 \begin{equation}
 H \! = \! \sum_{\br,\hat{\nu}}\left[\chi_{{\bf r}}^{\dag}\hat T_\nu\chi_{\br+\hat{\nu}} 
 + \mathrm{h.c.} \right]
+\sum_{\br}\chi_{\br}^{\dag}[V({\bf r})-m\sigma_z]\chi_{{\bf r}}
\label{eqn:ham}
 \end{equation}
where $\chi_r$ is a two-component spinor, $\hat T_\nu = t_{\nu}\sigma_z + t'_{\nu}\sigma_{\nu}$ is the usual kinetic energy hopping operator with strengths $t_{\nu}=t/2$ for $\nu=x,y,z$ and $t'_{\nu}=t'/2$ for $\nu=x,y$ and $t'_z=0$, $m$ controls the existence and location of the Weyl nodes, and $V(\mathbf r)$ is a random, on-site, potential (arising from disorder) drawn from a Gaussian distribution with zero mean and variance $W^2$
\begin{equation}
  \overline{V(\mathbf r)} = 0, \quad \overline{V(\mathbf r)V(\mathbf r')} = W^2 \delta_{\mathbf r \mathbf r'},
\end{equation}
where we denote disorder averaging by an over-line $\overline{(\cdots)}$.
This lattice model represents a time-reversal symmetry broken Weyl semimetal with four Weyl nodes for $|m|<t$, two Weyl nodes for $t< |m| < 3t$, and none (insulating) for $|m|>3t$.
Without disorder, the dispersion is $E_0(\bk) = \pm \sqrt{ t'^2[\sin(k_x)^2+\sin(k_y)^2]+[t\sum_{\nu}\cos(k_{\nu})-m]^2}$,
with Weyl points at $K_W=[0,0,\pm \arccos(m/t-2)]$.
We set $t=t'$ and 
$m = 3t/2$ so that we have two Weyl points at $K_W=(0,0,\pm2\pi/3)$ with one surface Fermi arc, 
an open boundary condition along $x$, and periodic boundary conditions along $y$ and $z$ (unless otherwise specified). 

We now first discuss the surface states in Eq.~\eqref{eqn:ham} 
without disorder $[V(\br)\equiv0]$ with a semi-infinite system $x \geq 1$.
With $k_y$ and $k_z$ as good quantum numbers, the effective 1D Hamiltonian is $H_0 = \sum_{x,k_y,k_z} H^{1\mathrm D}(x,\mathbf k_\perp)$ where $\mathbf k_\perp = (k_y, k_z)$ and 
\begin{align}
    H_{1D}=  \left( \tilde{\chi}_{x}^{\dag}\hat{t}  \tilde{\chi}^{\phantom{dag}}_{x+1} + \mathrm{H.c}\right)+  \tilde{\chi}_{x}^{\dag}\hat{\mu}  \tilde{\chi}_{x}^{\phantom{dag}}
\end{align} 
where $\hat{t}=(t\sigma_z+it'\sigma_{x})/2$ and $\hat{\mu}=[t(\cos k_y+\cos k_z -m )\sigma_z- t' \sin k_y\sigma_{y})]$.
Considering only a semi-infinite slab with $x>0$, general theory \cite{Mong2011} can then be used to find the surface states, which are usually written in terms of two exponentials $|\psi| \sim \lambda_1^x - \lambda_2^x$, but here we focus on the simple case $t=t'$ where $\lambda_2 = 0$.
This simple case has one solution exponentially decaying in $x$ such that the surface state wavefunction is given by
%$\psi_s(x,y,z) = e^{i(k_y y+k_z z)}f_S(x)\phi{\color{red}/L}$ 
%and has the surface dispersion $E_S$,
\begin{eqnarray}
\psi_S(x,y,z) &=& e^{i(k_y y+k_z z)}f_S(x)\phi/L,
 \\
f_S(x) &=& \sqrt{1-\lambda^{2}}\,  \lambda^{x-1}, \quad 
%\\
\end{eqnarray}
and has a surface dispersion %$E_S$,
\begin{eqnarray}
E_S(\mathbf k_\perp) &=&t \sin(k_y),
\end{eqnarray}
%\begin{equation}
%f_S(x)=\sqrt{1-\lambda^{2}}\,  \lambda^{x-1}, \quad
%E_S(\mathbf k_\perp) =t \sin(k_y),
%\end{equation}
%where $E_S$ denotes the surface dispersion, 
with a spinor $\phi^T=(1,-1)/\sqrt{2}$, and $\lambda = -([\cos(k_y)+\cos(k_z)] -m/t)$. 
The other surface (if the sample is instead finite along the $x$-direction) carries the opposite chirality with a dispersion $E_S=-t\sin(k_y)$. 
Valid solutions only exist for $|\lambda|<1$, defining the Fermi arc.
In Fig.~\ref{fig:dispersion-cuts}(c,d) we show some cuts through momentum space where the edge states are clearly identified.
The states are chiral (the group velocity $\mathbf v_g = \partial E_S/\partial \mathbf k_\perp$ is only nonzero along the $y$-direction). 
While these arcs are straight lines, our results presented here are independent of this feature (see Appendix~\ref{app:curvatureEffects}).

We first determine the bulk phase diagram at the Weyl node energy ($E=0$) as a function of disorder strength ($W$) by computing the average and typical density of states (DOS) using KPM with periodic boundary conditions in all directions. 
Following methods utilized in Refs.~\cite{Pixley2015,Pixley2016,Pixley2016a,JHWilson2017pip,Pixley2017}, we determine the location of the AQCP to be $W_c/t=0.9\pm0.025$, and the bulk Anderson localization transition at $W_l/t \approx 5.6-6.0$.
This gives us the bulk phase diagram in Fig.~\ref{fig:dispersion-cuts}(b).
Details and a short review of these methods are given in Appendix~\ref{app:Phase-Diagram}.

\begin{figure*}[h!]
\includegraphics{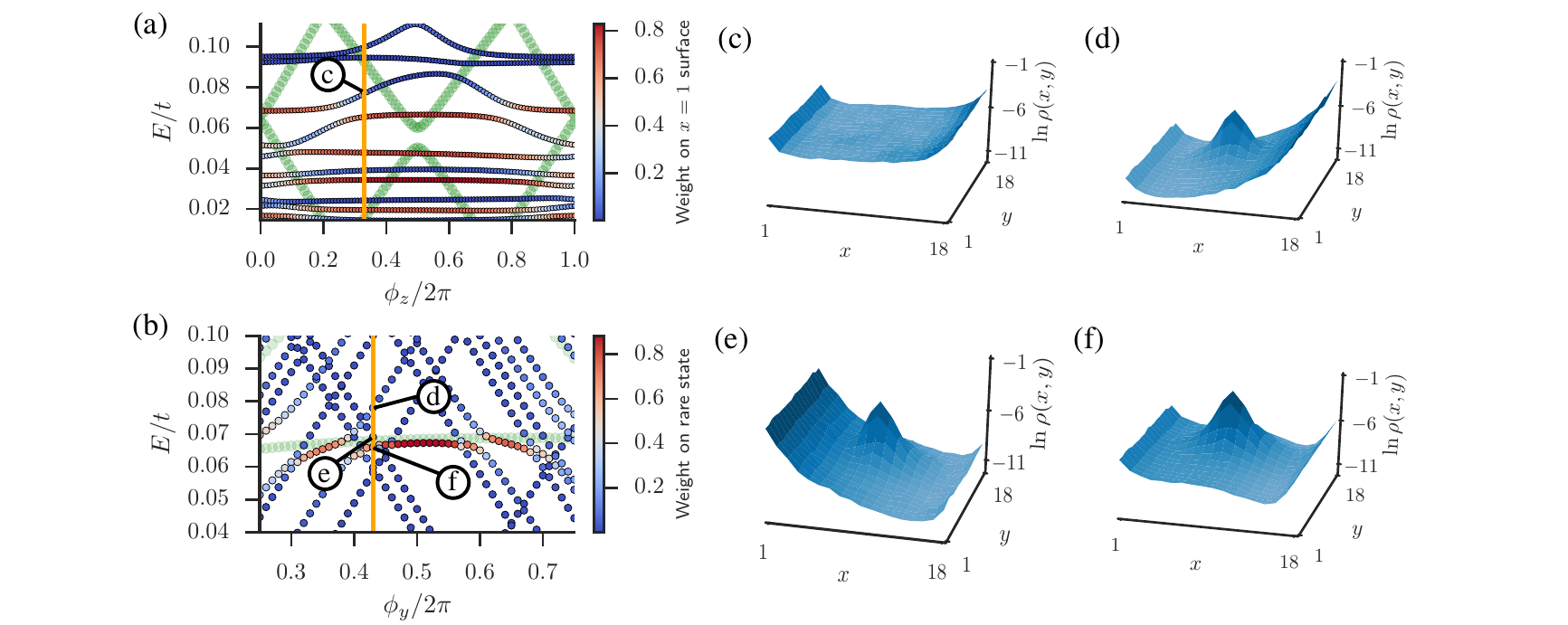}
\caption{(a) The low-energy eigenstates as a function of a twist in the $z$-direction for a disordered sample \emph{without} any rare states; total weight of the eigenstate on the $x=1$ surface is indicated by the color-scale. 
The indicated state shown in (c) represents hybridization between bulk Weyl states and surface states.
Green represents the bulk states found with periodic BCs.  (b) The low energy eigenstates as a function of a twist in the $y$-direction for a sample with a rare bulk state; The weight of the wave function on the rare state is indicated by the color scale; green again represents bulk states found with periodic BCs. Opposite chiral velocities represent states on opposing surfaces. The rare state hybridizes with both surfaces (d,e,f), strongly renormalizing the dispersion (b). The density plots are partially summed $\rho(x,y) = \sum_z |\braket{x,y,z|\psi}|^2$, and all plots are at weak disorder $W/t=0.5$ and have $L=18$.}
\label{fig:WF}
\end{figure*}

\section*{Spectral Features of the Arc}

For weak disorder, we can track the average arc states in momentum ${\bf k}$ despite ${\bf k}$ not being a good quantum number. 
To study the spectral features of the arc states (as probed in ARPES) we compute the disorder-averaged retarded Greens function on the surface $G(\br_i,\br_j,\omega)$; we focus on the relative distance between $\br_i$ and $\br_j$ and Fourier transform only on a surface $S(x)$ for $\mathbf r_i=(x,y_i,z_i)$ to $\mathbf k_\perp = (k_y, k_z)$
\begin{align}
  G^{\alpha\beta}(x,x';\mathbf k_\perp,\omega) = \braket{x,\mathbf k_\perp; \alpha| \tfrac1{\omega - H + i0^+} | x', \mathbf k_\perp; \beta}
\end{align}
with 
\begin{equation}
\ket{x, \mathbf k_\perp; \alpha} = \frac1{\sqrt{L^2}} \sum_{y,z}e^{i(k_yy+k_z z)}\chi_{\mathbf r,\alpha}^{\dag} \ket{0}
\end{equation}
(for the spinor component $\alpha$). Note that $|x,\mathbf k_\perp;\alpha\rangle$ is not an eigenstate of $H$ in the clean limit.
Focusing on the surface $x=x'=1$, the surface Green function and corresponding spectral function are given by
%$
\begin{eqnarray}
G_S(\mathbf k_\perp,\omega) &\equiv& \overline{ G(1,1;\mathbf k_\perp,\omega) },
\\
A_{S}(\bk_{\perp},\omega) &=& -\mathrm{Im}G_S(\bk_{\perp},\omega)/\pi,
\end{eqnarray}
%$,
which allow us to track properties of the arcs in momentum space.

As shown in the electronic dispersion curves of Fig.~\ref{fig:GreensfunctionTypicalDOS}(a), the features in the clean limit survive weak disorder but are broadened smoothly with increasing disorder, forming Fermi arc peaks in the surface spectral function $A_{S}(\bk_{\perp},\omega) $.
%$
%{\color {blue}
%\begin{equation}
%A_{S}(\bk_{\perp},\omega) = -\mathrm{Im}G_S(\bk_{\perp},\omega)/\pi .
%\end{equation}
%$.
%}
We also find finite energy bulk bands with weight on the surface that are well separated in energy from the surface states at weak disorder. 
Tracking the zero energy Fermi arc peak as a function of disorder [Fig.~\ref{fig:GreensfunctionTypicalDOS}(b)] shows that 
for weak disorder the Fermi arc peak at $\bk_{\perp}=0$ remains sharp and separate from the bulk states at finite energy. 
With increasing disorder, both the Fermi arc peaks and the bulk finite energy states on the surface broaden, which leads to the peak disappearing
around $W=1.0t$.  
This can be captured quantitatively with the width of the spectral function $\Gamma(\mathbf k_\perp) \equiv \mathrm{Im} 1/G_S(\mathbf k_\perp,\omega=0)$. 
As shown in the inset of Fig.~\ref{fig:GreensfunctionTypicalDOS}(b) (after converging in the KPM expansion order $N_C$ and finite size $L$, see Appendix~\ref{app:convergeWidthSpecFcn}), we find the Fermi arc peaks to smoothly broaden with increasing disorder and show no sign of the bulk crossover due to the AQCP. 
Therefore, in momentum space disordered surface and bulk are indistinguishable near the edge of the surface-bulk band, so
at moderate disorder strength, we must investigate a different observable.

The \emph{chiral} Fermi arc states propagate in one direction on each surface, Fig.~\ref{fig:dispersion-cuts}. 
Due to the absence of back-scattering, we expect weak disorder to not localize the surface states, but coupling to the bulk states complicates this picture.
To study the Anderson localization properties on the surface, we compute the typical DOS (i.e.\ the geometric mean of the local DOS) on the surface, defined by 
\begin{equation}
\rho_{t,S}(E)=\exp\left(\frac{1}{A_s}\sum_{i\in S(0)}^{A_s}\overline{\log\rho_i(E)}\right)
\end{equation}
where $A_s$ is a randomly chosen set of sites on the surface. 
At weak disorder we find the surface typical DOS approaching the average in the large-$L$ limit as seen in the inset to Fig.~\ref{fig:GreensfunctionTypicalDOS}(c), and thus the surface states are not localizing for small disorder, despite being two-dimensional. 
Further, the localization transition at large disorder ($W_l$) occurs in the bulk and on the surface simultaneously [see Fig.~\ref{fig:GreensfunctionTypicalDOS}(c) and Appendix~\ref{app:Phase-Diagram}].

\section*{Surface-Bulk Hybridization}
Thus far, we have not shown if the Fermi arc hybridizes with the bulk or if it is somehow ``protected.'' 
We first address these features on average explicitly by considering how a zero energy quasiparticle on the arc tunnels into the bulk.
We will primarily focus on the spectral weight associated with this process and therefore focus on 
%$
\begin{equation}
A_{||}(x,x',\mathbf k_\perp;\omega)=\frac{1}{\pi}\overline{|\mathrm{Im}G(x,x'; \mathbf k_\perp;\omega)|}
\end{equation}
%$ 
(we take the symmetric sum over $x$ and $x'$ and average the absolute value to suppress an average sign in the bulk). 
In the clean limit ( $W=0$)  and along the arc, the zero energy spectral function goes as 
$A_{||}(x,0,\mathbf k_\perp;\omega=0)\sim \exp[-x /\xi(\mathbf k_\perp)]$ 
(with the effect of the opposite surface being negligible), at the edge of the arc $\bk_{\perp}=(0,\pm2\pi/3)$, $\xi\rightarrow \infty$ and at $\mathbf k_\perp=\bm 0$, $\xi(\mathbf k_\perp)=\ln(2)$. 
This is shown in Fig.~\ref{fig:GreensfunctionTypicalDOS}(d) for three representative surface momenta on the arc $\bk_{\perp}=0$, at the Weyl node projection $\bk_{\perp}=(0,2\pi/3)$ and off the arc $\bk_{\perp}=(0,\pi)$ at weak disorder $W/t=0.5$. This shows that the two surfaces have become coupled on and off the arc.

We now determine the contribution of individual eigenstates to the average spectral function $A_{||}(x,x',\mathbf k_\perp;\omega)$.
To address this, we consider the low-energy properties using Lanczos on $H^2$.
Comparing with periodic boundary conditions shows that surface states are filling in the soft bulk gap, and twisting the boundary conditions reveals their chiral dispersion.
We first notice in Figs.~\ref{fig:WF}(a,c) that low-lying surface states hybridize weakly with bulk Weyl states. 
However, the hybridization between arc and bulk Weyl states vanishes at the Weyl energy in the limit of large-$L$, basically just due to the perturbative vanishing of the bulk DOS while the surface DOS remains nonzero.  Scattering to surface states near the Weyl nodes does perturbatively produce power-law tails in the bulk for the local DOS of the surface arc states at the Weyl energy~\cite{Gorbar2016}.

The surface arc states do hybridize with the non-perturbative rare bulk states. 
In contrast to TIs where the rare states are always exponentially bound in the gap, and therefore, cannot couple the two surfaces at arbitrary distances, these WSM rare states fall off as $1/r^2$ (see \cite{Nandkishore2014,Pixley2016,JHWilson2017pip} and Appendix~\ref{app:RareState}), and therefore a finite density of them couples the two surfaces at an arbitrary distance for any momenta.
As shown in Fig.~\ref{fig:WF}(b,d,e,f) we take such a rare bulk state found with periodic boundary conditions and then open the boundary well away from the location of the rare state. 
We find that this rare bulk state hybridizes with either surface (d,e) or even both surfaces (f) thus coupling the two surfaces and renormalizing the velocity of the chiral surface state, strongly reducing its magnitude [as $\partial E/\partial \phi_y$ in (b) depicts]. 
Therefore, we have shown that the arc states are not protected against disorder-induced hybridization with bulk rare states. 
Indeed, the rare states are spread out in momentum and have nonzero bulk DOS, so this non-perturbative surface-bulk hybridization occurs all along the arc and fully hybridizes in the large-$L$ limit with surface weight being $\sim 1/L$.
This non-zero density of the surface states deep in the bulk can be seen in Fig.~\ref{fig:GreensfunctionTypicalDOS}(d).

As we approach the cross-over to the metallic regime, many states begin to populate $E=0$ and will thus hybridize with the surface states. It is therefore in the semimetallic regime near $E=0$ (where the number of bulk Weyl states vanishes) that one might expect surface states to survive, but as we've shown, the existence of rare resonances within the bulk destroys even these.

\begin{figure*}[t]
\includegraphics[width=\columnwidth]{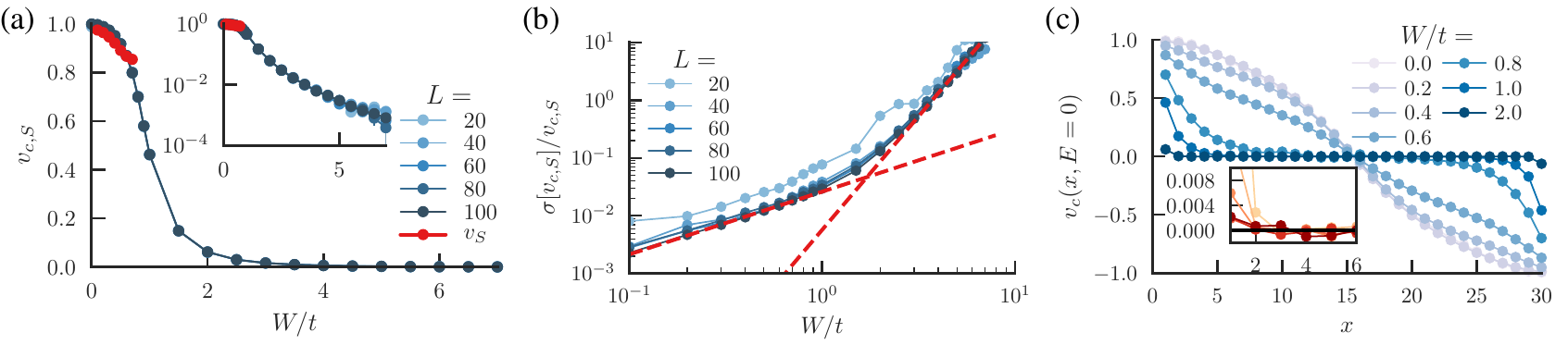}
\caption{(a) Average surface chiral velocity $v_c$ as a function of disorder on a linear and (inset) log scale for various system sizes. $v_S$ can be computed from the pole of the surface Green function only at weak disorder $W \le 0.6t$, we show $L=60$.  
(b) Broadness of the distribution of the surface velocity given by its standard deviation divided by the mean. The dashed lines are power law fits to the two distinct power law regimes, for $W\lesssim 1t$ we find $\sigma[v_{c,S}]/v_{c,s}\sim W^{1.08}$ and for $W> 1t$ it crosses over to $\sigma[v_{c,S}]/v_{c,s}\sim W^{3.9}$.
(c) Average velocity $v_c$ in each $y-z$ sheet located at $x$ for various disorder strengths. Disorder leads to a completely random velocity in the bulk but the surface chiral velocity persists to large disorder (inset) depicting disorder strengths $W=2.0t$ up to $7.0t$ in steps of $1.0t$.}
\label{fig:Jc}
\end{figure*}

\section*{Chiral velocity}

We find above that non-perturbative bulk rare states renormalize the chiral velocity of surface states, so the question arises: Can they drive the surface chiral velocity to zero?
To quantify this, we can study the dispersion as computed by the surface Green's function.
However, bulk states become an issue at finite disorder, filling in the pseudogap.
Therefore, we turn to a local measure of chiral velocity independent of the momentum, using a twist to define a layer-dependent velocity $v_c=\mathrm{Tr}_{S(x)}(\partial H /\partial \phi_y|_{\phi_y=0})$, where $\mathrm{Tr}_{S(x)}$ is a trace over the sheet at $x$; note that $J_y=-e\partial H /\partial \phi_y|_{\phi_y=0}$ is the current operator along the $y$-direction.
Using KPM, we project $J_y$ onto the sheet DOS at each energy and then divide by the sheet DOS to estimate the sum of matrix elements that contribute at that energy which yields the chiral \emph{velocity} at energy $E$ for sheet $x$
\begin{equation}
v_{c}(x,E)= \overline{\tfrac{\mathrm{Tr}_{S(x)}(J_y\delta(E-H))}{\mathrm{Tr}_{S(x)}(-e \delta(E-H))}}.
\label{eqn:Jc}
\end{equation}
We perform the trace stochastically after projecting onto each sheet $S(x)$. 
To study the zero energy average surface velocity we compute $v_{c,S}=(v_c(1,0)-v_c(L,0))/2$. 

As shown in Fig.~\ref{fig:Jc}(a), we find a very small finite size effect on the surface velocity, where on a linear scale $v_{c,S}$ appears to approach zero near the Anderson localization transition. 
However, when viewed on a log-scale [Fig.~\ref{fig:Jc}(a) inset], the data for $v_{c,S}$ is smooth through both the avoided transition and the localization transition; $v_{c,S}$ monotonically decreases for increasing $W$. 

Additionally, the distribution of the chiral velocity becomes increasingly broad for increasing $W$. 
To understand this, we can also characterize the statistics of this object on a per-sample basis with its variation
\begin{align}
    \sigma[v_{c,S}]^2 = \overline{\left(\tfrac{\mathrm{Tr}_{S(x)}(J_y\delta(E-H))}{\mathrm{Tr}_{S(x)}(-e \delta(E-H))} - v_{c,S}(x)\right)^2}.
\end{align}
The broadness of Eq.~\eqref{eqn:Jc} can be characterized with $\sigma[v_{c,S}]/v_{c,S}$ as seen in Fig.~\ref{fig:Jc}(b). We find the distribution becomes increasing broad as the model passes through the localization transition; the data for $\sigma[v_{c,S}]/v_{c,s}$ has two different power law regimes, for $W\lesssim 1t$ we find $\sigma[v_{c,S}]/v_{c,s}\sim W^{1.08}$ and for $W> 1t$ it crosses over to $\sigma[v_{c,S}]/v_{c,s}\sim W^{3.9}$ with a smooth evolution and no signature of the localization transition.

In Fig.~\ref{fig:Jc}(c) we show the velocity as a function of the distance along the system from each surface. For increasing disorder we find that the velocity in the middle of the system becomes completely random (and averages to zero) while the current on the two surfaces survives up to large disorder.
It is striking that we find a small but non-zero chiral velocity on the surface even inside the Anderson insulating phase.

\begin{figure*}
    \centering
    \includegraphics[width=\columnwidth]{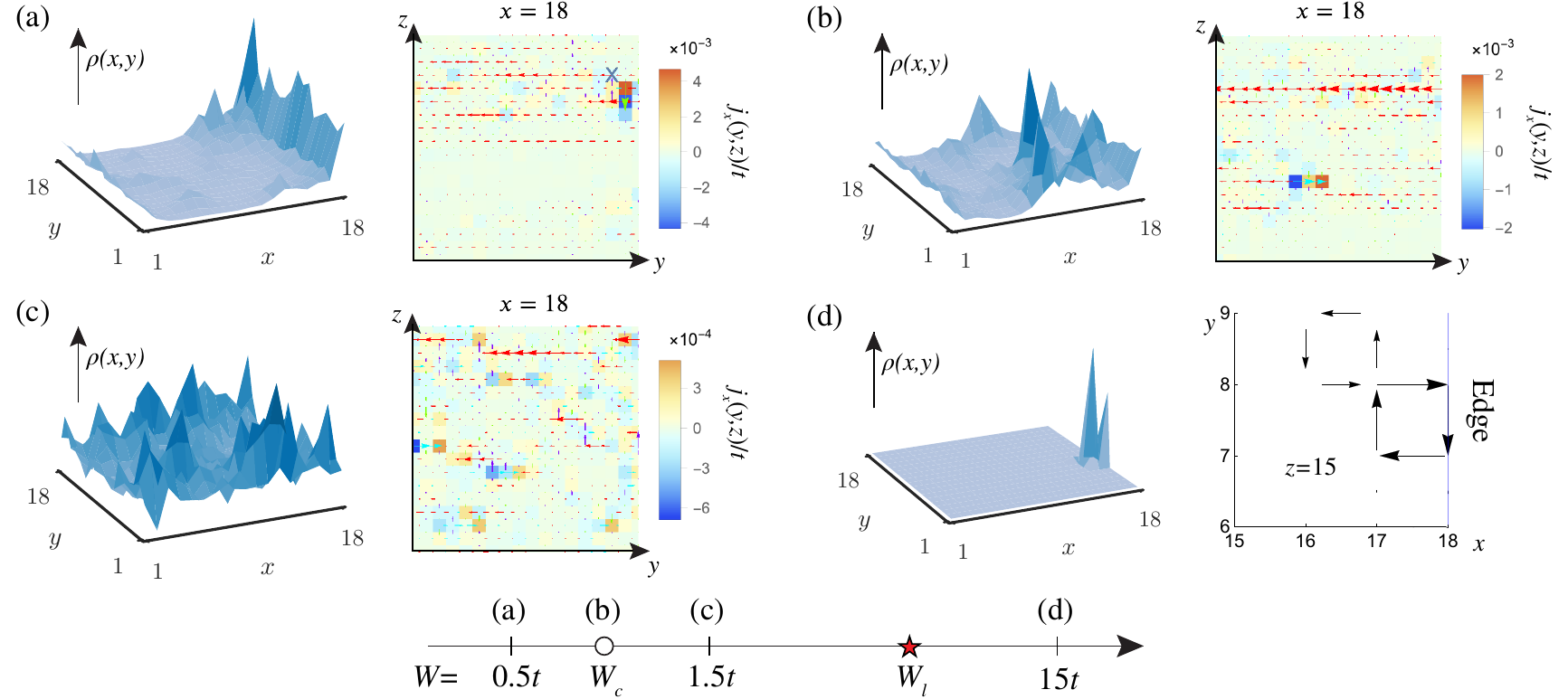}
    \caption{Plots of characteristic near-surface eigenfunctions: Their bulk density profile (left) and surface current (right). The density profile is partially summed $\rho(x,y) = \sum_z |\braket{x,y,z|\psi}|^2$, and the surface is at $x=L=18$. The current density profiles are normalized by the largest current value on any bond, with colored squares indicating current flowing onto or out of the surface (i.e.\ the surface divergence of the current). In (a) a rare state is located and moved close to the surface (indicated by a blue X in the current density profile). Notice that each wave function has more chiral velocity (red, left moving) on the surface, even when the state is delocalized throughout the bulk (c). The fully localized state in (d) has a surface chiral velocity but only as small current loop near the surface (and only near the plane $z=15$).}
    \label{fig:surfaceCurrents}
\end{figure*}

To address these features, we look at typical wave functions and the current along each bond in each of the regimes of the model in Fig.~\ref{fig:surfaceCurrents}.
First, in Fig.~\ref{fig:surfaceCurrents}(a) we see that at $W=0.5t$
the surface state is largely intact (though it is hybridized slightly with a rare state indicated by the blue X), and the surface %has a 
chiral velocity is largely intact. 
As we increase disorder to be roughly at the AQCP [Fig.~\ref{fig:surfaceCurrents}(b)], 
the current is still largely flowing in one chiral direction on the surface.
The same situation applies deep in the diffusive metal regime ($W=1.5t$) as seen in Fig.~\ref{fig:surfaceCurrents}(c) with a state that is exclusively a bulk state, but still hosts a chiral velocity on the surface.
Last, well beyond the localization transition ($W=15t$) [see Fig.~\ref{fig:surfaceCurrents}(d)]  
we clearly see a localized state near the surface, 
with a current loop  with chiral velocity that resides within the localization length.
Thus, in this regime the  currents loops are localized and will not produce a finite Hall conductivity, consistent with Refs.~\cite{chen_disorder_2015,shapourian_phase_2016}.
In this way, the system can simultaneously be fully localized and still have a preference for chiral velocity on the surface.

\section*{Conclusions}
To conclude, we have investigated the non-perturbative disorder effects on the surface states of a Weyl semimetal. 
The surface quasi-particles 
acquire finite lifetime and renormalized chiral velocity, but become ill-defined at moderate disorder strengths.
Our results on the surface spectral function demonstrate how the surface Fermi arcs can be observed in ARPES experiments without being topologically protected. 
We have established that rare non-perturbative bulk states hybridize with the Weyl Fermi arcs making them no longer bound to the surface even at aribtrarily weak disorder.
Nonetheless, 
we find that the surface chiral velocity 
persists to
quite 
large disorder strengths (independent of the amount of curvature along the arc), even past where the surface and bulk states Anderson-localize, by forming localized current loops while retaining their chiral nature on the surface.
Strikingly, this feature of the surface states persists despite the destruction of the sharp distinction between surface and bulk states and the disappearance of the WSM phase itself due to disorder.

\begin{acknowledgments} 
\emph{Acknowledgments---}We thank Pallab Goswami, Mehdi Kargarian, Rahul Nandkishore, and Jay Sau for useful discussions.
This work was performed in part at the Aspen Center for Physics (J.~P.\ and G.~R.), which is supported by National Science Foundation grant PHY-1607611.
The authors are grateful for support from the Air Force Office for Scientific Research (J.~W.), the Laboratory for Physical Sciences (J.~P. and S.~D.-S.), the Packard Foundation (G.~R.), and the IQIM an NSF-PFC (G.~R.).
The authors acknowledge the University of Maryland supercomputing resources (http://hpcc.umd.edu), the Beowulf cluster at the Department of Physics and Astronomy of Rutgers University, The State University of New Jersey, and the Office of Advanced Research Computing (OARC) at Rutgers, The State University of New Jersey (http://oarc.rutgers.edu) for providing access to the Amarel cluster and associated research computing resources that have contributed to the results reported here. 
\end{acknowledgments}

\appendix

\section{Effects of curvature to the chiral velocity}
\label{app:curvatureEffects}

\begin{figure}
    \centering
    \includegraphics{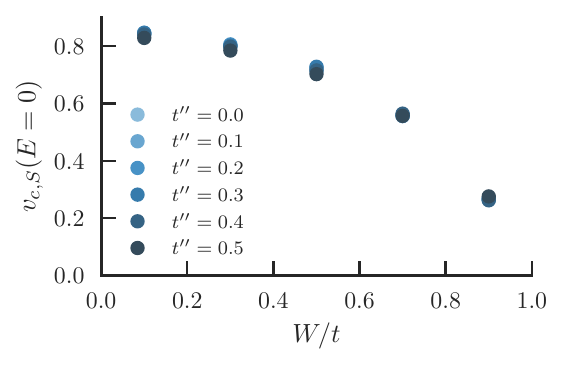}
    \caption{Tracking the chiral velocity at $E=0$, we see that a bend characterized by $t''$ does not affect the surface chiral velocity.
    These are results on a system of size $L=10$.}
    \label{fig:chiralvelocitywithbend}
\end{figure}

To add curvature to the arc, we add in an additional hopping term to the Hamiltonian
\begin{align}
    \Delta H = \frac{t''}2\sum_{\br}\chi_{{\bf r}}^{\dag} \sigma_y\chi_{\br+\hat{z}} 
 + \mathrm{h.c.},
\end{align}
which modifies our effective 1D Hamiltonian so that
\begin{align}
    \Delta H_{1D} = t'' \cos k_z\, \sigma_y.
\end{align}
The surface-localized wave functions are not affected by this change, but the dispersion changes
\begin{align}
    E_S(k_y,k_z) = t'\sin k_y + t'' \cos k_z.
\end{align}
The Fermi-arcs are no longer straight, but curved.
To test if this appreciably affects the results, we define the chiral velocity as the velocity perpendicular to the line intersecting the ends of the Fermi-arc (so it is still in the $y$-direction). 
Then, testing on small system sizes ($L=10$), we find, as seen in Fig.~\ref{fig:chiralvelocitywithbend} that as disorder is increased, the chiral surface velocity is relatively unaffected by $t''$.

\section{Phase Diagram}
\label{app:Phase-Diagram}

Using periodic BCs with the Hamiltonian \eqref{eqn:ham}, we establish the phase diagram for the bulk
\begin{center}
\includegraphics[width=0.4\columnwidth]{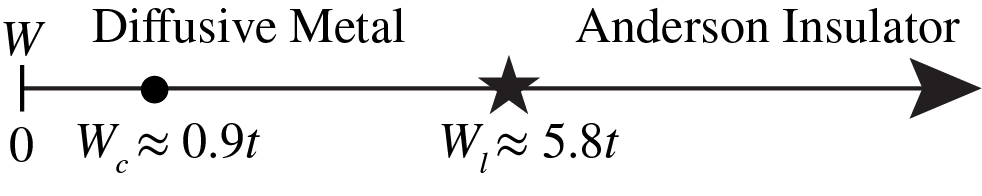}
\end{center}

\subsection{Avoided Quantum Critical Point, $W_c$}

To characterize the bulk phases we use the density of states defined for a system of size $L$ as
\begin{align}
    \rho(E) = \overline{\frac1{L^3}\sum_n \delta(E - E_n)},
\end{align}
where $E_n$ are the energies of the eigenstates of the system, and the overline $\overline{(\cdots)}$ represents disorder averaging.

Using the KPM, which we refer the reader to the the review \cite{Weisse2006} and previous works \cite{Pixley2016,Pixley2016a,JHWilson2017pip,Pixley2017}, we can numerically calculate the density of states (and other quantities) for large system sizes.
This method introduces a new finite size in the form of a series truncation, controlled by the variable $N_C$.
Balancing $N_C$ and $L$ are crucial to handling finite size effects appropriately.

\begin{figure}
    \centering
    \includegraphics{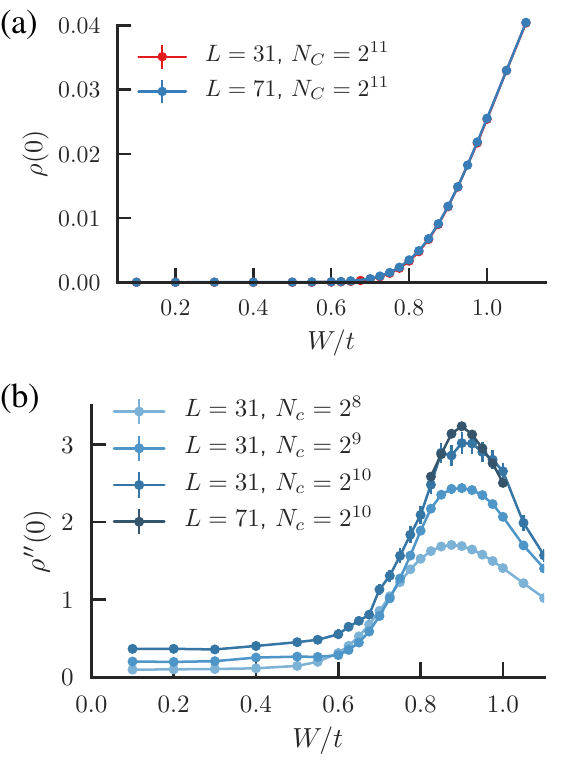}
    \caption{(a) The density of states at $E=0$ vs.\ disorder strength $W$ of this system as found from via the KPM method and (b) the saturation of the second derivative of the density of states $\rho''(0)$ vs.\ W. The peak of the former characterizes the location of the avoided quantum critical point.}
    \label{fig:BulkDOS}
\end{figure}

We are interested in the effects near $E=0$ where the semimetallic nature of the material is strongest.
We show $\rho(0)$ vs.\ $W$ in Fig.~\ref{fig:BulkDOS}.
Avoided critically is captured by the maximum of $\rho''(0)$ where for each $N_C$ we saturate $\rho''(0)$ in $L$ before moving to larger $N_C$.
Iterating this, we can converge a peak to $\rho''(0)$ as indicated in Fig.~\ref{fig:BulkDOS} and obtain $W_c/t = 0.900\pm 0.025$.

\subsection{Anderson Localization Critical Point, $W_l$}
\label{subapp:Wl}

Using methods similar to \cite{Pixley2015}, we can roughly estimate the location of the localization transition. 
To probe this, we can look at the local density of states (for site $i$ and realization $r$)
\begin{align}
\rho_{i,r}(E) = \sum_n |\braket{i|\psi_{n,r}}|^2 \delta(E - E_{n,r})
\end{align}
where $E_{n,r}$ and $\psi_{n,r}$ are respectively the energy and wave function for the $n$th eigenstate of the $r$th realization.
From this, we can define the typical density of states as the geometric average of this quantity
\begin{align}
    \rho_t(E) = \exp\left\{\overline{\frac1{L^3}\sum_i \log[\rho_{i,r}(E)]}\right\}.
\end{align}
Instead of a sum over all sites, in practice we take a random set of sites to average over.
The vanishing of this quantity is associated with the onset of localization.

\begin{figure}
    \centering
    \includegraphics{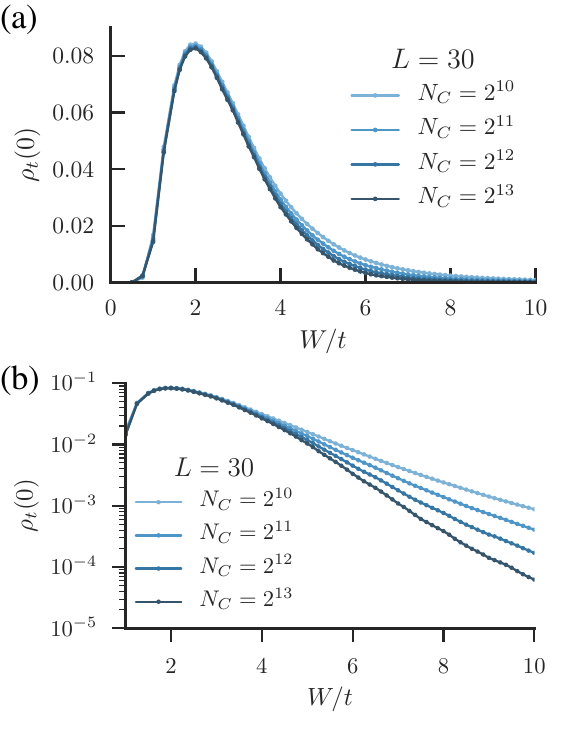}
    \caption{Plot of the typical density of states at $E=0$, $\rho_t(0)$. (a) Shows the decrease in the typical density of states which mirrors that seen in Fig.~\ref{fig:GreensfunctionTypicalDOS}(c). (b) Further, notice that dependence on $N_C$ begins to set in around $W_l/t \approx 5.6-6.0$ .}
    \label{fig:rhot0}
\end{figure}

With the KPM method though, the typical density of states does not formally vanish since the series cutoff $N_C$ smears out the wave functions.
Therefore, the typical density of states should begin to decrease with increased $N_C$ around the localization transition \cite{Pixley2015,JHWilson2017pip} as we see around $W\sim6.0t$ in Fig.~\ref{fig:rhot0}. 

\begin{figure}
  \centering
  \includegraphics{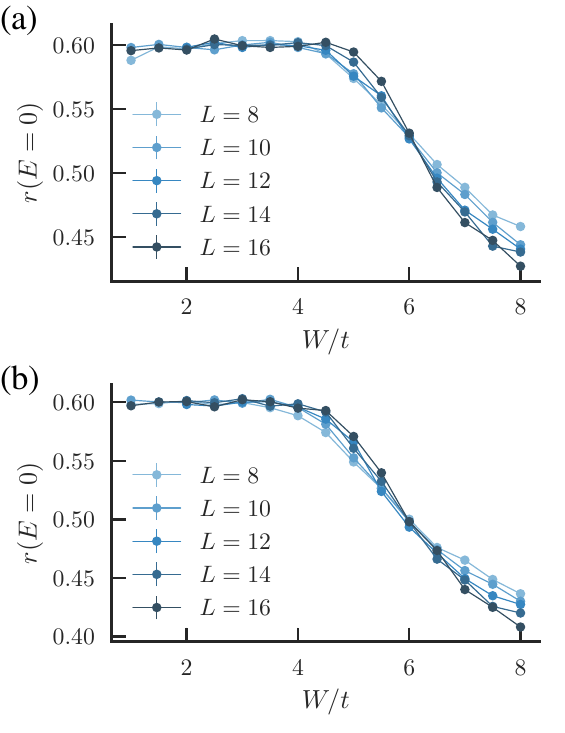}
  \caption{(a) The level statistics as they change for higher disorder with periodic boundary conditions. Notice the crossing roughly around $W\approx 6.0t$. (b) The level statistics as they change for higher disorder with \emph{open} bondary conditions. Notice the crossing roughly around $W\approx 6.0t$. The bins in energy space to determine both plots are roughly $4\%$ of the total bandwidth and symmetric about $E=0$. \label{fig:levelstats}}
\end{figure}

To get an estimate of the localization transition, we use the adjacent gap ratio on smaller system sizes
\begin{align}
  r_n = \frac{\min(E_{n+1}-E_n,E_n-E_{n-1})}{\max(E_{n+1}-E_n,E_n-E_{n-1})},
\end{align}
and we take the average of $r_n$ around a particular energy to produce $r = \overline{r_n}$.
Previous work shows that $r = 0.60$ for GUE (diffusive phase) and $r=0.386$ for a Poisson spectrum (localized phase) \cite{dalessio2014}.
We see $r$ change in Fig.~\ref{fig:levelstats} where we compare $r(E=0)$ (the value of $r$ around $E=0$) with periodic [left figure in Fig.~\ref{fig:levelstats}] or open [right figure in Fig.~\ref{fig:levelstats}] boundary conditions. We use $5 \times $ Freedman-Diaconis to bin eigenstates around $E=0$ to determine $r(E=0)$, and take 10-100 realizations.
From this data, we estimate that $W_l \approx 6.0t$ in rough agreement with what we see in Fig.~\ref{fig:rhot0}.

\section{Convergence of the width of the surface spectral function}
\label{app:convergeWidthSpecFcn}

\begin{figure}
    \centering
    \includegraphics{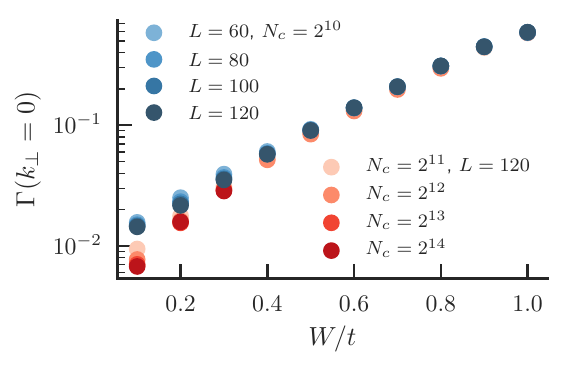}
    \caption{Saturation of the width of the peak of the surface spectral function  in $\omega$ defined as $\Gamma({\bf k}_{\perp})=1/\mathrm{Im}G_S({\bf k}_{\perp},\omega=0)$  with the surface wave-vector $\mathbf k_\perp = 0$. We are able to converge our results for $W\ge 0.1t$. }
    \label{fig:GF_WidthSaturation}
\end{figure}

In the main text, we present the converged width of the surface spectral function defined as $\Gamma({\bf k}_{\perp})=1/\mathrm{Im}G_S({\bf k}_{\perp},\omega=0)$ for ${\bf k}_{\perp}=0$.
%make use of the surface Green's function, but it is important to make sure our results are converged in the finite sizes $L$ and $N_C$.
%To see how results converge, we consider the width of the peak at surface wave-vector $\mathbf k_\perp = 0$ in frequency space.
This peak is associated with zero energy Weyl Fermi arc surface states, and the
width $\Gamma({\bf k}_{\perp})=1/\mathrm{Im}G_S({\bf k}_{\perp},\omega=0)$ continuously increases with increasing disorder strength. The dependence of the peak width on finite system size $L$ and expansion order $N_C$ is  shown in Fig.~\ref{fig:GF_WidthSaturation}. To make sure the peak is not artificially broadened we follow the same procduer as in Ref.~\cite{Pixley2017}. We first shift the random potential to sum to zero for each disorder sample (this eliminates the leading finite size effect from perturbative effects~\cite{Pixley2016}). To eliminate finite size effects we work at $N_C=2^{10}$ and vary $L$ until the data is roughly $L$ independent at $L=120$. We then fix $L=120$ and vary $N_C$ until the peak is independent of both $L$ and $N_C$. Applying this procedure we can converge the width of the peak for disorder strengths $W\ge0.1t$.

\section{Characterizing the rare state wave function}
\label{app:RareState}

To study the rare state's effect on surface states, we had to isolate a rare state with a system that has periodic boundary conditions, then open them to see how it hybridizes with surface states.

\begin{figure*}
    \centering
    \includegraphics{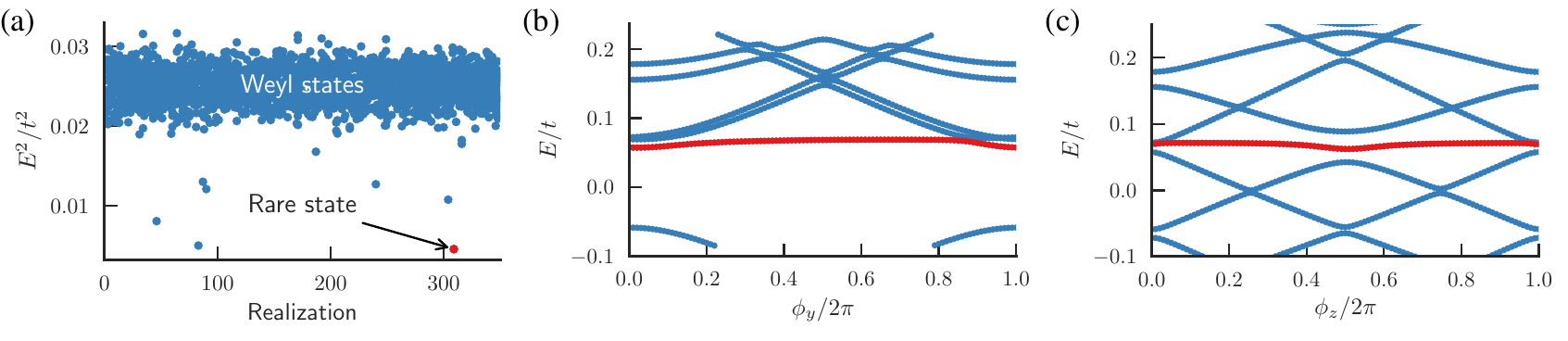}
    \caption{(a) Twisting the boundary conditions moves the bulk Weyl states away from zero energy, revealing low-lying rare states; here we focus on the lowest one pictured, $r=309$. It has energy $E \approx 0.067t$ and remains stable when boundary conditions are twisted (b, c). This data is for a disorder strength $W=0.5t$ and $L=18$.}
    \label{fig:identifying_rare_state}
\end{figure*}

Working with $L=18$, we first maximally move the bulk Weyl states away from zero energy with a twist in the boundary conditions.
Running a number of realizations as shown in Fig.~\ref{fig:identifying_rare_state}(a) we pick out a potential candidate for a rare state.
Here it is realization $r=309$.
We can then twist the boundary conditions to see that this is indeed a rare state that does not respond appreciably to the twisted boundary conditions [see Fig.~\ref{fig:identifying_rare_state}(b,c)].

\begin{figure}
    \centering
    \includegraphics{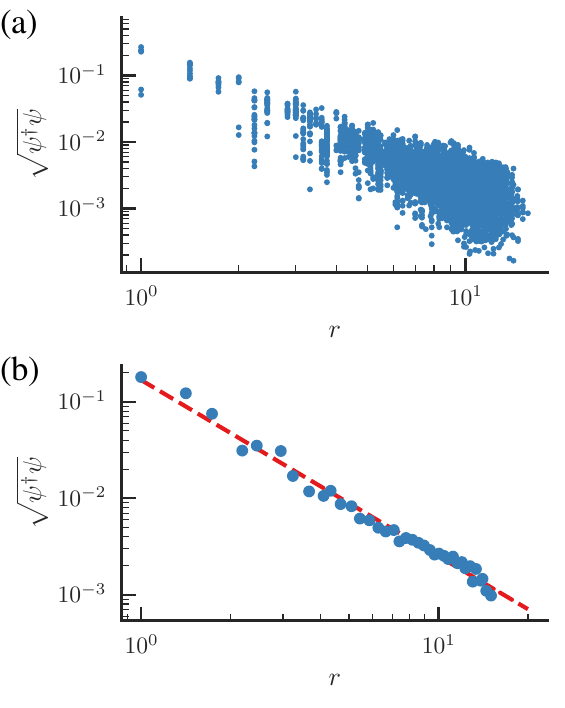}
    \caption{(a) The absolute size of the wave function on sites a distance $r = |\mathbf r - \mathbf r_{\mathrm{max}}|$ from the maximum. (b) The result of binning the wave function and the red-line is a power law fit to the resulting data: $1/r^{1.83}$.}
    \label{fig:rarestate-distance}
\end{figure}

To determine how localized it is, we find the maximum of the wavefunction at $\mathbf r_{\mathrm{max}}$, then determine how the wave function falls off as a function of radius $r = |\mathbf r - \mathbf r_{\mathrm{max}}|$.
We bin the data using the Freedman-Diaconis rule, and we then fit a power-law to the resulting binned data (see Fig.~\ref{fig:rarestate-distance}).
The result is a power law (red line on the right figure of Fig.~\ref{fig:rarestate-distance}) of
\begin{align}
    |\psi(\mathbf r)| \sim \frac1{r^{1.83}},
\end{align}
which is consistent with the analytic prediction of a power law of $1/r^2$. This state is found to hybridize wth surface states as we see elaborate on in the main text.

% Bibliography
\bibliography{extracted}

\end{document}